\documentclass[reprint,flushbottom,showpacs]{revtex4-1}

\usepackage{psfrag} 
\usepackage{graphicx}

\usepackage{amsmath}
\usepackage{amssymb}
\usepackage{amsthm} 
\usepackage{bm}
\usepackage{color}

\newcommand{\ket}[1]{\left|#1\right>}
\newcommand{\bra}[1]{\left<#1\right|}
\newcommand{\nn}{\nonumber\\}

\newcommand{\bea}{\begin{eqnarray}}
\newcommand{\ea}{\end{eqnarray}}
\newcommand{\eea}{\end{eqnarray}}
\newcommand{\ord}{{\cal O}}
\newcommand{\sumint}[1]

\newcommand{\tfix}{t_{\mathrm{fix}}}
\newcommand{\Tfix}{T_{\mathrm{fix}}}
\DeclareMathOperator\erf{erf}

\begin{document}

\title{Stability of spherically trapped three-dimensional Bose-Einstein condensates\\ against macroscopic fragmentation}

\author{Philipp Bader$^{1}$ and Uwe R. Fischer$^{2}$}

\affiliation{$^1$Universitat Polit\`ecnica de Val\`encia, Instituto de Matem\'atica Multidisciplinar,  
E-46022 Valencia, Spain\\
$^2$Seoul National University,   Department of Physics and Astronomy \\  Center for Theoretical Physics, 
151-747 Seoul, Korea}

\begin{abstract}
We consider spherically trapped Bose gases in three dimensions with contact interactions,  and investigate whether the Bose-Einstein condensate at zero temperature is stable against macroscopic fragmentation into a small number of mutually incoherent pieces. Our results are 
expressed in terms of a dimensionless interaction measure proportional to the Thomas-Fermi parameter.  
It is shown that while three-dimensional condensates are inherently much more stable against macroscopic 
fragmentation than their quasi-one- and quasi-two-dimensional counterparts, they fragment at a sufficiently 
large value of the dimensionless interaction measure, which we determine both fully numerically and semianalytically from 
a continuum limit of large particle numbers.   
\end{abstract}


\maketitle

\section{Introduction}
Bose-Einstein condensation \cite{Bose1924,Einstein1924} of noninteracting bosons  
can in principle occur in arbitrarily large spatial dimension $D$, depending both on the 
properties of the single-particle spectrum and the confining 
potential \cite{Groot}.  
On the other hand, the Hohenberg-Mermin-Wagner theorem rules out, independent of the strength of interactions, Bose-Einstein
condensation in $D\le 2$ \cite{Hohenberg,MerminWagner}. 
The latter theorem, however, applies to homogeneous condensates in the thermodynamic limit,
where long-range phase fluctuations trigger the decay of the Bose-Einstein condensate into infinitely many fragments. 
This changes for trapped condensates, where the finite extension of the gas cuts off
the phase fluctuations in the corresponding directions. While an explicitly interaction-independent formulation of the theorem is still possible, the geometric shape of the condensate enters the Bogoliubov inequality on which the Hohenberg-Mermin-Wagner theorem rests \cite{Fischer}.

It is well known that in three spatial dimensions, fragmentation does not occur in the thermodynamic limit and in a homogeneous system for positive interaction coupling \cite{Nozieres,Mueller}, while 
for negative coupling constant the system is unstable.
For a trapped, that is spatially localized and inhomogeneous system, in three dimensions, the relevant dimensionless parameter to measure the importance of interactions over the single-particle kinetic and trapping contributions to the energy is the Thomas-Fermi parameter $Na_s/l_0$ (where $a_s$ and $l_0$ are $s$-wave scattering and harmonic trapping length, respectively).
It was previously observed by us that harmonic trapping and positive interaction coupling can lead 
to fragmentation into two mutually incoherent macroscopic pieces, 
forming a so-called fragmented condensate, 
well before the thermodynamic limit is taken for quasi-one-dimensional (quasi-1D) and quasi-two-dimensional
(quasi-2D) gases  \cite{Bader,FischerBader}. 
To more completely elucidate the dimension dependence of the many-body physics of fragmentation, 
we present here a detailed analysis of fragmentation for the completely symmetric example a spherically trapped 3D condensate. 
Viewed from a different angle, we investigate to which extent the conventional textbook wisdom \cite{PethickSmith,Dalfovo}, that when the Thomas-Fermi parameter of a 3D spherically trapped 
condensate 
is going to infinity yields a interaction-dominated single 
Bose-Einstein condensate (with parabolic shape in this Thomas-Fermi limit and in a harmonic trap) 
needs revision. 

In three dimensions, due to the spherical symmetry of the  system, even when the field operator expansion is restricted to the low-energy sector, there are potentially four single-particle states which are macroscopically
occupied. By numerical analysis and general symmetry arguments, 
we find that fragmentation is dominated by two orbitals
at a dimensionless coupling measure which is proportional
to the Thomas-Fermi parameter. The critical coupling measure  
is one (two) orders of magnitude larger than the corresponding measure in the quasi-2D (quasi-1D) cases. 
In addition, the maximal degree of fragmentation \cite{Bader} turns out to be significantly smaller than in the latter cases. 
Our result therefore implies the rapidly growing persistence of an interacting, trapped scalar Bose-Einstein condensate against macroscopic fragmentation 
upon increasing the spatial dimension.  

\section{Spherically trapped gases in three dimensions}
\subsection{The four-mode approximation for the Hamiltonian}

To facilitate comparison with the previously treated quasi-1D and quasi-2D trapping cases, 
we will make a one-parameter variational ansatz for the single-particle orbitals as in \cite{Fischer}.
This involves ground and first excited states of the harmonic oscillator, 
with the variational parameter chosen to be harmonic oscillator length.  
Compared to fully self-consistent multiconfigurational Hartree calculations as performed, e.g., in \cite{SAC,ASC,SSAC,Lu}, 
while being less quantitatively accurate, the variational approach 
leads to a qualitatively correct picture of the fragmentation phenomenon.
A particular merit of this approach is that the parameter dependence of the fragmentation transition is transparent:  
We find that fragmentation is decided by a single parameter, $G_3$ in Eq.\,\eqref{defG3} below, 
which measures the relative
importance of interactions over the single-particle (trapping) energies.  
In addition, the variational approach is capable to deal with the limit of very large particle numbers $N$; 
in its continuum limit, which we will derive below, there is indeed no upper bound to the value of $N$. 
This is particularly beneficial in three spatial dimensions, where the particle numbers at the same densities are obviously larger than in one- and two-dimensional systems; we were able to numerically 
calculate within relatively short timescales systems with up to 
$N\sim 10^6$ particles.
 
 
To formulate the proper variational orbitals basis, we first write down the well-known eigenstates and energies of 
the isotropic harmonic oscillator in three spatial dimensions, 
\bea
\psi_{nlm} & = & Y_l^m (\theta,\phi) N_{nl} 
r^l 
\exp\left[- \frac{r^2}2\right]   
L_n^{(l+1/2)} \left(r^2\right),
\nn 
E_{nl} &=& \omega \left(2n+l+\frac32\right), \label{ho3D}
\ea
with normalization $N_{n,l} = \frac{\sqrt{\Gamma(n+l+\frac12)}}{\sqrt{n!}\Gamma (l+\frac12)}$, and 
the functions $Y_l^m,L_n^{(l+1/2)}$ are spherical harmonics and generalized Laguerre polynomials
respectively. 

{The width of the single-particle basis functions is expressed by a length $R$, which is the scaling length
of the radial coordinate $r$, that is we put $r/R \rightarrow r$.}
For {\em noninteracting} condensates the width $R$ is given by the harmonic oscillator length, 
$R=l_0=\omega^{-1/2}$ ($\hbar = M  =1$, where $M$ is the boson mass). In the following, we assume 
$R$ to be a variational parameter, which will
determine the family of solutions of the many-body equations, i.e., 
whether single or fragmented condensates are obtained is determined by variation of $R$. {In order to make large particle number calculations feasible, as well as to render the energy landscape of the eigenvalue problem in its most transparent form, the dimensionality of the variational space is reduced by assuming that all single-particle orbitals scale with the same $R$.}

According to \eqref{ho3D}, the four energetically lowest states are given by the quantum numbers 
$n=0,\,l=0$ (ground) and $n=0,\,l=1,\,m=-1,0,1$ (first excited)  
[adopting the Condon-Shortley phase convention for spherical harmonics], 
\bea
\psi_0 &\equiv& \psi_{000}  =  \frac1{\pi^{3/4}}  \exp\left[- \frac{r^2}2\right], \nn 
\psi_1 &\equiv& \psi_{010}  =  \sqrt2 r\cos\theta \psi_0(r),\nn
\psi_+ &\equiv& \psi_{011} =  -r e^{i\phi}\sin\theta\psi_0(r),\nn
\psi_- &\equiv& \psi_{01-1} =  r e^{-i\phi}\sin\theta\psi_0(r). \label{four modes} 
\ea
{The contact-interaction many-body Hamiltonian reads, written in terms of the full field operators,  
\bea 
\label{H in field form}
\begin{split}
	\hat{H} &=
	\int \mathrm{d}^3 {x} ~~
	\hat{\Psi}^{\dag} ({\bm x})
	\bigg[
	-
	\frac{ \nabla^2}
	{2}
	+ 	\frac12 \omega^2 r^2 
	\bigg]
	\hat{\Psi} ({\bm x}) 
	\\
 &	+\frac{g}{2}
	\iint \mathrm{d}^3 {x} ~~
	\hat{\Psi}^{\dag} ({\bm x})
	\hat{\Psi}^{\dag} ({\bm x})
	\hat{\Psi} ({\bm x})
	\hat{\Psi} ({\bm x}).
\end{split}
\ea
After} truncating the field 
operator expansion, including the four modes \eqref{four modes}, we obtain 
\bea
\hat H &=& \sum_{i=0,1,\pm} \left[ \epsilon_i \hat n_i 
+\frac12  C_i \hat n_i (\hat n_i-1) \right] \nn 
& & +\frac12 D_1 \hat n_0 \hat n_1
+\frac12 D_2 \left(\hat n_0 \hat n_{+} + \hat n_0 \hat n_{-} \right) \nn 
& & 
+\frac12 D_3 \left(\hat n_1 \hat n_{+} + \hat n_1 \hat n_{-}\right)
+\frac12 D_4 \hat n_{+}\hat n_{-}
\nn & & +\frac12 \left\{
E_1 \hat a^\dagger_1 \hat a^\dagger_1 \hat a_0 \hat a_0 
+E_2 \hat a^\dagger_+ \hat a^\dagger_- \hat a_0 \hat a_0 
+ E_3  \hat a^\dagger_+ \hat a^\dagger_- \hat a_1 \hat a_1 
\right\} \nn & & +{\rm h.c.} 
\label{4mH} 
\ea 
Note that the pair-exchange scattering (terms $\propto E_i$) 
occurs also between energetically degenerate orbitals 
(degenerate on the single-particle level), being represented by the term
$\propto E_3$, and involving the excited states $m=\pm 1$ and $m=0$ ($l=1$). This is 
distinct from the quasi-1D and quasi-2D cases treated in \cite{FischerBader}, 
where pair-exchange scattering only occurs between pairs of ground and excited single-particle states.

The interaction matrix elements 
$V_{ijkl} = g \int\int\int r^2 \sin\theta dr d\theta d\phi\,
\psi^*_i ({\bm r})\psi^*_j ({\bm r}) \psi_k ({\bm r})\psi_l ({\bm r})$
are related to the coefficients in \eqref{4mH} as follows.
The non-vanishing pair-exchange matrix elements are 
\bea
E_1 &=&V_{1100} ,\qquad
E_2=V_{+-00} + V_{-+00},  
\nn
E_3&=& V_{+-11} + V_{-+11}. 
\ea
The remaining coefficients are of the density-density type, 
\bea
C_0 &=& V_{0000},\qquad C_{1}=V_{1111},\qquad 
C_{+} =V_{++++},\nn 
C_{-}&=&V_{----}, \qquad 
D_1 =  V_{0101}+V_{1010}+V_{1001}+V_{0110} , \nn 
D_2 &=& D_1 (1\rightarrow \pm),\qquad 
D_3 = D_1(0\rightarrow \pm ), \nn
D_4 &=& D_1(0\rightarrow +,1\rightarrow -).
\ea
The result for the scattering coefficients (reinstating now the variational 
harmonic oscillator length), may be written in a compact notation in the following way,  
\begin{multline}
\{C_0,C_1,C_\pm,D_1,D_2,D_3,D_4,E_1,E_2,E_3\} \\
= \frac g{(2\pi)^{3/2}R^3}\left\{1,\frac34,\frac1{2},
2,4,2,2,
\frac12,1,\frac{1}{2}\right\}.
\label{eq:scatter_coeff}
\end{multline}
The single-particle energies are given by  
$\epsilon_i = \int r^2\sin\theta dr d\theta d\phi
\left[|\nabla\psi_i|^2/2 +\omega^2 r^2|\psi_i|^2/2\right]$, 
and read 
\bea
\epsilon_0 &=&  \frac34\left[\frac1{R^2} +\omega^2R^2 \right],\quad
\epsilon_1 =   \frac54\left[\frac1{R^2} +\omega^2R^2 \right],\nn
\epsilon_\pm &=&   \frac54\left[\frac1{R^2} +\omega^2R^2 \right] =\epsilon_1.
\ea
Defining the scaled variational parameter $\Lambda = {R}/{l_0}$, we have as the typical units 
of single-particle energies and coupling constants
\bea
\epsilon_0 =\frac34\omega \left(\frac1{\Lambda^2}+\Lambda^2\right), \qquad
C_0 = \frac{G_3\omega}{N\Lambda^3}, 
\ea
where we introduced the  dimensionless interaction coupling
\bea 
G_3 = \frac{Ng}{(2\pi)^{3/2}l_0}. \label{defG3} 
\ea 
Like its quasi-1D and quasi-2D counterparts $G_1=\frac{Ngl_z}{(2\pi)^{3/2}l_\perp^2}$ 
and $G_2=\frac{Ng}{(2\pi)^{3/2} l_z}$, where $l_z$ and $l_\perp$ are harmonic oscillator lengths of
a cylindrical trap, the quantity $G_3$  measures the relative
importance of total interaction and kinetic and potential energy terms in the Hamiltonian.
In the present spherically trapped 3D case, $G_3$ is simply directly proportional to the 
well-known Thomas-Fermi parameter \cite{Dalfovo}, which, as will be demonstrated below, is the 
single parameter deciding the question of coherence versus fragmentation. 

\subsection{Wavefunction ansatz and the eigenvalue problem}

We employ a general four-mode ansatz in the Fock subspace of fixed total particle number
\begin{align}
\ket{\Psi} &= \sum_{l_1,l_\pm} \psi_{l_1,\,l_+,\,l_-}\ket{N-l_1-l_+-l_-,\,l_1,\,l_+,\,l_-}\!.
\end{align}
The total energy $E=\bra\Psi \hat H \ket\Psi$ 
in terms of the level occupation amplitudes $\psi_{l_1,l_+,l_-}$ then reads
\begin{widetext}
\begin{multline}
 E =
c_{l_1,l_\pm}\sum_{i=0,1,\pm} |\psi_{l_1,l_\pm}|^2 
+\frac12 E_1 d_{1} \psi^*_{l_1,l_+,l_-} \psi_{l_1+2,l_+,l_-} 
 +\frac12 E_2 d_{2} \psi^*_{l_1,l_+,l_-} \psi_{l_1,l_++1,l_-+1}  
+\frac12 E_3 d_{3} \psi^*_{l_1,l_+,l_-} \psi_{l_1-2,l_++1,l_-+1} \\
 +\frac12 E_1 d_{1} \psi^*_{l_1+2,l_+,l_-} \psi_{l_1,l_+,l_-} 
+\frac12 E_2 d_{2} \psi^*_{l_1,l_++1,l_-+1} \psi_{l_1,l_+,l_-} 
   +\frac12 E_3 d_{3} \psi^*_{l_1-2,l_++1,l_-+1} \psi_{l_1,l_+,l_-},  
   \label{Energy}
   \end{multline}
where the diagonal and pair-exchange coefficients take the explicit form  
\bea
c_{l_1,l_\pm} &=& 
\epsilon_0 \left(N- \sum_{i=1,\pm}
\right) + \sum_{i=1,\pm} \epsilon_i l_i 
+\frac12 C_0 \left(N- \sum_{i=1,\pm}l_i\right) 
\left(N- \sum_{i=1,\pm}l_i
-1\right)
+\frac12 \sum_{i=1,\pm} C_i l_i(l_i-1)
\nn 
& & 	+\frac12 D_1 \left(N- \sum_{i=1,\pm}l_i 
	\right) l_1
+\frac12 D_2\left(N- \sum_{i=1,\pm}l_i 
\right)\left(l_++l_-\right)
+\frac12 D_3 l_1(l_++l_-) +\frac12 D_4 l_+ l_-, \nn
d_1(l_1,l_+,l_-) &=& \sqrt{\left(N- \sum_{i=1,\pm}l_i 
-1\right)\left(N- \sum_{i=1,\pm}l_i 
\right)\left(l_1+2\right)\left(l_1+1\right)},\\
d_2(l_1,l_+,l_-) &=& \sqrt{\left(N- \sum_{i=1,\pm}l_i 
\right)\left(N- \sum_{i=1,\pm}l_i 
\right)\left(l_++1\right)\left(l_-+1\right)},\quad \nonumber
d_3(l_1,l_+,l_-) = \sqrt{l_1(l_1-1)(l_++1)(l_-+1)}.
\ea
Finally, the minimization of the energy functional \eqref{Energy} with respect to  $\psi^*_{l_1,l_+,l_-}$ 
gives the eigenequations 
\bea \label{eq:full_eigenv_problem}
E\psi_{l_1,l_+,l_-} &=& c_{l_1,l_\pm} \psi_{l_1,l_+,l_-} 
+ \frac{E_1}2 d_1 (l_1,l_+,l_-) \psi_{l_1+2,l_+,l_-} 
+ \frac{E_2}2 d_2 (l_1,l_+,l_-) \psi_{l_1,l_++1,l_-+1} \nn 
& & + \frac{E_3}2 d_3 (l_1,l_+,l_-) \psi_{l_1-2,l_++1,l_-+1} 
+ \frac{E_1}2 d_1 (l_1-2,l_+,l_-) \psi_{l_1-2,l_+,l_-} 
+ \frac{E_2}2 d_2 (l_1,l_+-1,l_--1) \psi_{l_1,l_+-1,l_--1} \nn
& & + \frac{E_3}2 d_3 (l_1+2,l_+-1,l_--1) \psi_{l_1+2,l_+-1,l_--1} .
\ea
\end{widetext} 
\section{Solving the eigenvalue problem}

\subsection{Decomposition into smaller problems: The $k$-subspaces}
We are facing a high-dimensional eigenvalue problem that is difficult to solve for typical particle numbers because the matrix dimensions scale roughly with $N^3\times N^3$ when na\"ively implemented. However, the assumed orbitals allow for algebraic simplifications, to be explained in what follows, in order to significantly reduce the problem size.
{We note in this context that in \cite{Tsatsos}, a thorough analysis of angular momentum many-body states has been performed for attractively interacting and rotating Bose gases.}

Observe that only a particular set of couplings between the $l_1$ and $l_\pm$ terms appear in eq. \eqref{eq:full_eigenv_problem}. To be more specific, only couplings between terms where $l_+-l_-=\text{constant}$ are allowed as a consequence of the model.
Similar to the reduction for a three-mode model in the quasi-2D case treated in \cite{FischerBader}, this allows us to partition the eigenvalue problem into $2N+1$ smaller problems, by introducing the notation $\psi^k_{l_1,l_+}\equiv\psi_{l_1,l_+,l_-}$ with $-N\leq k\equiv l_--l_+\leq N$ and $\psi=0$ for indices such that $|l_1+2l_++k|>N$.
In terms of the many-body amplitudes with index $k$, the new eigenvalue problem, with the relations we have found for the matrix elements, c.f. \eqref{eq:scatter_coeff}, reads
\begin{widetext}
\bea
\label{eq:eigenv_problem_me}
E\psi^k_{l_1,l_+} &=& c^k_{l_1,l_+} \psi^k_{l_1,l_+} 
+ \frac{E_1}2 d_1^k (l_1,l_+) \psi^k_{l_1+2,l_+} 
+ \frac{E_1}2 d_1^k (l_1-2,l_+) \psi^k_{l_1-2,l_+} 
+ \frac{2E_1}2 d_2^k (l_1,l_+) \psi^k_{l_1,l_++1}  \nn& & 
+ \frac{2E_1}2 d_2^k (l_1,l_+-1) \psi^k_{l_1,l_+-1} 
+ \frac{E_1}2 d_3^k (l_1,l_+) \psi^k_{l_1-2,l_++1} 
+ \frac{E_1}2 d_3^k (l_1+2,l_+-1) \psi^k_{l_1+2,l_+-1}.
\ea
\end{widetext}
Since $\epsilon_-=\epsilon_+$ and $C_+=C_-$, we have $c_{l_1,l_+,l_-}=c_{l_1,l_-,l_+}$ and $d_1,d_2,d_3$ are also symmetric with respect to interchanging $l_+$ and $l_-$. 


The eigenvalue problem \eqref{eq:full_eigenv_problem} then becomes completely symmetric in $l_+$ and $l_-$, i.e., interchanging them leaves the equation unaltered and we can assume them, backed up with numerical simulations, to be identical for the many-body ground state, up to a global phase $\phi$ in the amplitudes, that is 
$$
		l_+ \equiv l_- 
	\;\text{or formally}\;
		\forall\, l_1,l_\pm: \psi_{l_1,l_+,l_-}=e^{i\phi}\psi_{l_1,l_-,l_+}.
$$
{Physically, this implies an overall zero-angular momentum for the ground state, as expected for our fully symmetric setup with repulsive interaction.}


We can immediately deduce from \eqref{eq:eigenv_problem_me} that the ground state at $k=0$ will be (nearly) degenerate due to the decoupling of even and odd values of $l_1$. The problem can thus be split further to separate the practically degenerate eigenstates (up to the energy of one particle) that belong to only even or only odd occupation numbers $l_1$.
These substantial size reductions allow to solve the eigenvalue problem numerically and yields a matrix size of approximately $N^2\times N^2$.  
Due to the coupling structure of the Hamiltonian, 
this matrix is very sparse and the total number of nonzero entries grows only quadratically with the particle number $\mathcal{O}(N^2)$.

We remark that the even-odd degeneracy 
allows for the free choice of a phase parameter $\theta$ 
in the superposition of the degenerate eigenstates \cite{Kangsoo}, which depends on the preparation of the 
state; $\theta\equiv 0$ in what follows. 



\subsection{Numerical results}
The ground states for different configurations $N,G_{3}$ have been computed numerically by finding local minima in the energy curve along the variational parameter $\Lambda$. 

Numerical calculations for particle numbers up to $N=20000$ confirm that $k>0$ states correspond to increasingly higher energies, with the lowest difference ($k=0$ to $k=1$) in energy 
per particle being approximately $\ord{(G_3^{0.44}/N \omega)}$ and relative energy differences $1-E_{k=1}/E_{k=0}$ also scaling with $\mathcal{O}(1/N)$. 

\begin{figure}\centering
	  \psfrag{LEFTXLABEL}[][l]{\rule{16mm}{0mm}\textsf{max. $l_\pm$}}
	  \psfrag{LEFTYLABEL}[][c]{log$_{10}|$ Error in energy $|$}
	  \psfrag{LEFTTITLE}[][c] {Influence of higher $l_\pm$ states}
	  \psfrag{RIGHTXLABEL}[][l]{\rule{18mm}{0mm}\textsf{$l_\pm$}}
	  \psfrag{RIGHTYLABEL}[][c]{\textsf{$l_1$}}
	  \psfrag{RIGHTTITLE}[][c]{$N = 5000,\; G_3 = 5000$}
	\includegraphics[width=0.495\textwidth]{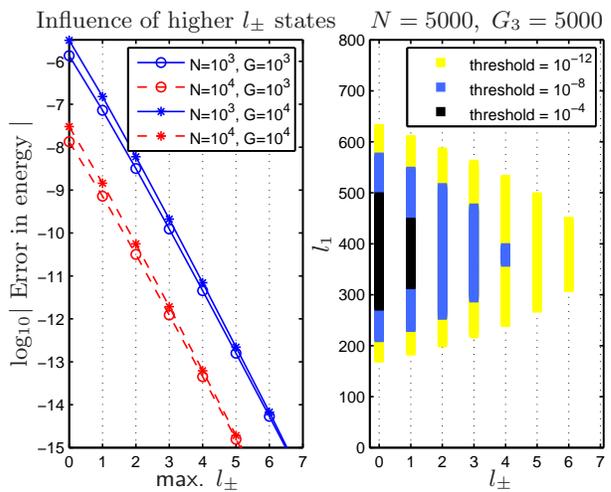}
	\caption{\label{fig:small_lp}(color online) 
	The left panel shows the error committed by truncation of the eigenvalue problem at a fixed value $\max l_\pm$ for different values of the particle number $N$ and interaction strength $G_3$ with
	 $\Lambda_\mathrm{min}$ fixed at the minimal variational energy configuration. Curves with same line style and symbol correspond to same $N$ or $G_3$, respectively. The right plot visualizes the locations of amplitudes $|\psi_{l_1,l_+}^0|^2$ larger than a given threshold. Note the scaling on the $l_1$ axis which is of order $\mathcal{O}(N)$, whereas the horizontal axis ends at $l_\pm=7$.
	 }
\end{figure}

Fixing $k$ at zero, and hence $l_+ =l_-$, 
the occupation of the circulating orbitals stays $\sum_{l_1,(l_\pm>0)}|\psi_{l_1,l\pm}|^2<3\%$, with all significant amplitudes located at $l_\pm \sim \mathcal{O}(1)$, whereas the occupation of the radially symmetric orbital at $l_\pm=0$ is scaling with $l_1\sim\mathcal{O}(N)$, cf. Fig.~\ref{fig:small_lp}.
Including as few as eight circular states is sufficient to reach machine accuracy for all reasonable configurations $N,G_3$.

For large values of $G_3$ a second shallow minimum appears in the energy landscape, analogous to the quasi-1D and quasi-2D cases \cite{FischerBader}, and the condensate starts to fragment. 
We have determined numerically that the onset of fragmentation is determined  by a critical value of 
the interaction parameter, $(G_3)_c$, which depends on particle number. For small $N\sim 1000$, the critical interaction strength is $(G_3)_c\approx 5600$, a value that decreases quickly to its asymptotic value $(G_3)_c=2480\pm10$ for $N=50000$. 
For large values of $G_3$, the fragmented local minimum becomes a global one and the non-fragmented minimum becomes very shallow.

\begin{figure}\centering
	  \psfrag{YLABEL}[][c]%
	  	{\textsf{Fragm. }$\mathcal{F}$, \textsf{ 1 - weight in }$l_\pm$=0}    
	  \psfrag{XLABEL}[][c]%
	  	{$N$}
	  \psfrag{TITLE}[][c]%
	  	{$G_3 = 5000$}	  	
		\psfrag{YLABEL}[][c]%
	  	{Fragmentation $\mathcal{F}$}
\includegraphics[width=.495\textwidth]{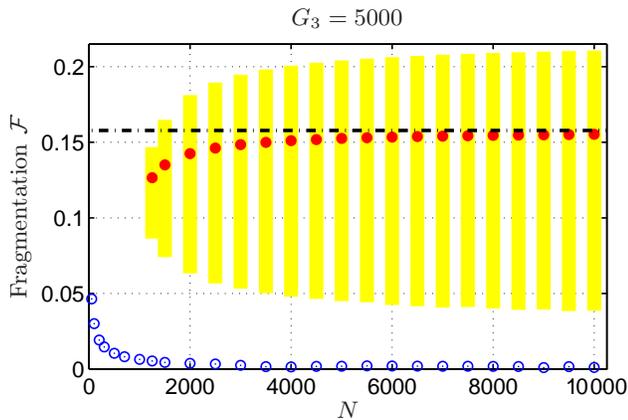}
		\caption{\label{fig:FoverN} (color online)
		Degree of fragmentation (circles)
			 of the ground state for varying $N$ at fixed $G_3=5000$ for nonfragmented (blue empty circles) and fragmented 
			 (red filled circles) states. Small $N$ effects include a nonvanishing fragmentation which quickly approaches $0$. 
			After passing a critical $N$, a new minimum appears and asymptotes its maximum for moderate values of $N$.
			The dashed-dotted black line shows the degree of fragmentation at $\mathcal{F}=0.16$, computed in the large $N$ limit \eqref{eq:continuum:energy}. 
			Vertical bars (yellow) indicate the sensitivity of the degree of fragmentation when we allow for an energy variation away from the local minimum via  $\Lambda$, up to the energy barrier separating the two minima.
				}
\end{figure}	
		
A finite particle number effect on the fragmentation can be observed by varying $N$ for given $G_3$ and is illustrated in Figure~\ref{fig:FoverN}. For small particle numbers, corresponding to a subcritical $G_3(N)$, only one minimum exists and its degree of fragmentation, defined by
$\mathcal F = 1-|\lambda_1-\lambda_2|/N$, where $\lambda_i$ are the (macroscopic) eigenvalues of the single-particle density matrix \cite{Bader}, rapidly approaches zero when $N$ is increased.
Once we have passed the critical value for $N$, fragmentation sets in, with the appearance of a new local energy minimum at a smaller extension $\Lambda$, and quickly approaches a limit which depends on the chosen value of $G_3$. 

\begin{figure}\centering
	  \psfrag{XLABEL}[][c]%
	  	{$\log_{10} G_3$}	  	
	  \psfrag{TITLE}[][c]%
	  	{\textsf{$N = 20000$}}	  	
		\psfrag{YLABEL}[][c]%
	  	{Fragmentation $\mathcal{F}$}
		\includegraphics[width=.495\textwidth]{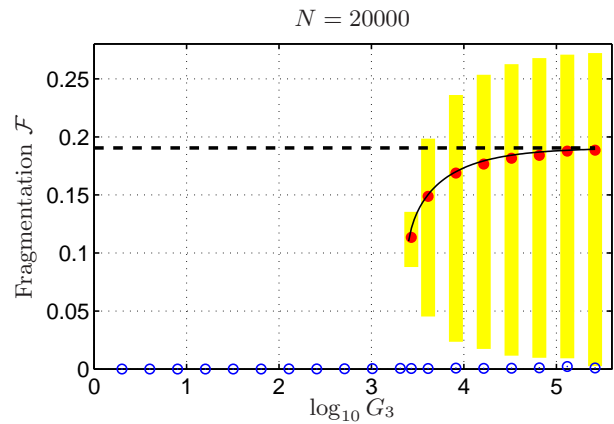}
		\caption{\label{fig:FoverG} (color online) Dependence of degree of fragmentation (circles) 
		on $G_3$ for fixed $N=20000$. The single existing minimum (blue empty circles) for subcritical $G_3$ shows no fragmentation, whereas the one created after a critical $G_3$ has been passed (red filled circles) soon asymptotes to its limiting value $\mathcal{F}=0.19$ indicated by the dashed line. The solid line corresponds to the continuum limit results computed via the minimization of \eqref{eq:continuum:energy}.}
\end{figure}

The dependence of the fragmentation
on the interaction strength $G_3$ is depicted in Fig.~\ref{fig:FoverG}. Note the onset of fragmentation after passing the critical $(G_3)_c(N) = \ord{(10^3)}$. 

The considerations above on the low occupancy of circulating states, cf. Fig.~\ref{fig:small_lp}, allow to simplify the problem by neglecting the small $l_\pm>0$ contribution. This yields a 
two-mode model, for which we apply a continuum limit \cite{Spekkens} in the following section.

\section{The continuum limit of the effective two-mode model}
\subsection{Derivation of the Schr\"odinger equation  for the mode population}
From the full numerical analysis, we are led to conclude that $l_\pm\neq0$ is approximately unpopulated and after deletion of the $l_\pm>0$ terms, the eigenvalue problem \eqref{eq:eigenv_problem_me} reduces to
\begin{multline} \label{eq:eigenv:conti}
E\psi^k_{l_1,0} = c_{l_1,0} \psi^k_{l_1,0}\\
	+ \frac{E_1}2 d_1 (l_1,0) \psi^k_{l_1+2,0}
	+ \frac{E_1}2 d_1 (l_1-2,0) \psi^k_{l_1-2,0} .
\end{multline}
The approximations (omitting the subscript $1$)
$d_1(l) \approx \tilde d_1 \equiv -(l - N/2
)^2 + N^2/4$ and $d_1(l+2)\approx d_1(l)$ 
yield
$
	d_1(l)\left(\psi_{l+2}-2\psi_l+\psi_{l-2}\right) \approx 4d_1(l)\partial_l^2,
$
which we use to write \eqref{eq:eigenv:conti}
as an ordinary differential equation
\begin{equation}
	4\frac{E_1}{2} \tilde d_1(l) \partial_l^2 \psi(l) + c(l)+ 2\frac{E_1}{2}d_1(l) \psi(l) = E\psi(l).
\end{equation}
We employ a change of variables, neglecting from here on $\ord(1/N)$ terms, 
$t = l-N/2
$, and obtain
\begin{multline}
	 2E_1\left[-t^2 + N^2/4\right] \partial_t^2 \Psi(t) + \\
	 \left[E_1(-t^2 + N^2/4)  + c(t+N/2
	 ) \right] \Psi(t) = E\Psi(t).
\end{multline}
We identify the above equation with the Schr\"odinger equation of the harmonic oscillator via
\begin{align}
-\frac{1}{2m}\partial_t^2 \Psi(t) + 
 \left( \frac12 m \omega \left( t - \mathfrak{S}\right)^2 + e_{\mathfrak{S}} \right)\Psi(t) 
  = E\Psi(t), \nn \label{HO}
 \end{align}
with the parameters
\begin{equation} 
\begin{split}
	m 		 &= \frac{1}{-4E_1\left(N^2/4 - \tfix^2 \right)},\\
	\omega &= \sqrt{- E_1(C_0 + C_1 - D_1 - 2 E_1) (N^2 - 4 \tfix^2)},\\
	\mathfrak{S} 		 &= \frac{(\epsilon_0-\epsilon_1) + (C_0-C_1)(N-1)/2}{C_0 + C_1 - D_1 - 2 E_1},\\
	e_{\mathfrak{S}} &= E_1 \frac{N^2}{4} + c_{N/2} - \frac12 m\omega^2\mathfrak{S}^2,
\end{split}  
\end{equation} 
and an energy shift $e_{\mathfrak{S}}$ independent of $t$.
Note that we have fixed the contribution from the mode-exchange ($\propto E_1$) at $t=\tfix$, that is taking $d_1(\tfix)$ instead of $d_1(t)$, in front of the derivative. 
With the scalings
$t=T\cdot N/2$, $\tfix=\Tfix N/2$ and $E_1=\frac12 C_0$, we then have $\mathfrak{S}=\frac N6 \left(1-\frac{16}{3}X\right)$
in terms of the ratio of single-particle energy to interaction energy units $X=\epsilon_0/(NC_0)$.

The ground state of \eqref{HO} can be solved for analytically when the absolute value of the Fock-state amplitudes $|\psi(t)|$ is considered as a continuous variable \cite{Spekkens,Bader},
\bea\label{eq:psi(t)}
	|\psi(t)| = \frac{1}{\left(\pi \sigma^2\right)^{1/4}} \exp\left[-\frac{ \left( t-\mathfrak{S}\right)^2}{2 \sigma^2}\right].
\ea
We get the effective oscillator length of the ``harmonic oscillator''
(note that $T\in[-1,1]$) as 
\bea
\sigma^2 = \sqrt{\frac1{m\omega}} = N\sqrt{\frac23(1 - \Tfix^2)}.
\ea
The single-particle to interaction energy units ratio $X$ is then calculated to be  
\bea
	X &=& \frac{\epsilon_0}{NC_0} 
	= \frac34 \frac{1}{G_3}\left(\Lambda+\Lambda^5\right).
\ea
Finally, the total energy in the continuum limit for the reduced model is given by
\begin{align}
	E = \omega + e_{\mathfrak{S}}= \omega + E_1 \frac{N^2}{4} + c_{N/2} - \frac12 m\omega^2\mathfrak{S}^2.
\end{align}
We note that, to this order, the dependence of $\sigma$ on $\Tfix$ does not enter the continuum energy; 
we finally obtain, to first order in $N$,
\bea
	\frac{E}{NC_0} = \frac{N}{3} + \frac{13N}{9}X - \frac{8N}{27}X^2 + \mathcal{O}\left(1\right), 
\ea
Then, with $C_0={G_3\omega}/({N\Lambda^3})$,
\bea\label{eq:continuum:energy}
	\frac{E}{N\omega} = 
							\frac{13}{12}\left(\frac1{\Lambda^2}+\Lambda^2\right)
						- \frac{\Lambda^3}{6G_3} \left(\frac1{\Lambda^2}+\Lambda^2\right)^2
						+ \frac{G_3}{3\Lambda^3}, \nn 
\ea
which represents the continuum expression for the energy as a function of $\Lambda$, with the sole parameter $G_3$. 


\subsection{Large coupling limit}
The minimization problem $\partial E/\partial \Lambda =0$ in the limit of \eqref{eq:continuum:energy} can be solved for real values of $G_3$ when $\Lambda>4.686$, and we get for the minimum 
\bea
	G_{3} &=& 
	\frac{13}{12} (\Lambda^5-\Lambda) + \frac{1}{12} \sqrt{193 \Lambda^2 - 482 \Lambda^6 + \Lambda^{10}}\nn
	&\approx& \frac{7}{6}\,\Lambda^5 - \frac{127}{6} \Lambda +\mathcal{O}(\frac{1}{\Lambda^3}) \approx \frac{7}{6}\,\Lambda^5.
\ea 

For the relative energy difference to a Fock state with all particles occupying the $l=0$ state (the radial
ground state), we get 
\bea
	\frac{\Delta E}{N\omega} = \frac{E_{\text{Fock}} - E_{\text{Cont.}}}{N\omega} \sim 
	0.00013\, G_3^{2/5}, 
\ea
with the Fock state energy
\bea
\frac{E_{\text{Fock}}}{N\omega}=\frac34(1/\Lambda^2+\Lambda^2)+\frac{G_{3}}{2\Lambda^3}.
\ea
At its minimum, $\Lambda_{\text{Fock,min}}^5\approx G_3$. 
Note also that in quasi-1D, we had 
$\Delta E/(N\omega)\sim 0.02 G_1^{2/3}$ and in quasi-2D, 
$\Delta E/(N\omega_\perp)\sim 0.002 G_2^{1/2}$ \cite{Fischer}. Hence, with increasing dimension, both the prefactor
as well as the scaling of the energy difference to a single condensate decrease.  
For a second, fragmented minimum to exist we need $G_3>2436.13$, and then have
$\Delta E/(N\omega) \gtrsim 0.0015$.

The continuum limit is valid around the expansion point $\tfix$, which we put equal to the shift, $\Tfix\equiv\mathfrak{S}/(N/2)=\frac13-\frac{16}9 X$.
For the single-particle to interaction-energy ratio, we have
$
X
=\frac{3}{4}\left(\frac{6}{7}\right)^{1/5} G_3^{-4/5} + \frac{9}{14}
$, 
which asymptotes to $X=\frac9{14}\approx 0.64$ and is close to this value already for the critical $(G_3)_c=2436.13$.
Then, the asymptotic shift is evaluated to $\mathfrak{S}/(N/2)\approx-0.810$, giving 
$\sigma^2 = N \sqrt{\frac23 (1-\Tfix^2)}\approx 0.48 N.$
We can now assess the validity of the continuum approach by measuring the occupation it assigns to negative (unphysical) $l_1$.
With increasing $N$, the width of the wavefunction \eqref{eq:psi(t)} gets smaller, and the density at negative $l_1$ goes to zero as $\frac12(1-\erf[0.13729 \sqrt{N}])$. The rapid convergence is illustrated for $N=1000$, when $\int_{-\infty}^{-N/2}|\psi(t)|^2 \,\mathrm{d}t\approx 4.1\cdot 10^{-10}$.

\subsection{Degree of fragmentation}
The degree of fragmentation in the continuum limit reads 
\bea
{\cal F} &=&
1- \frac 2N \sqrt{\left[\frac N2 \sin\theta \left(1-\frac{\sigma^2 +
2{\mathfrak S}^2}{N^2}\right)\right]^2\!+{\mathfrak S}^2} .\nn
\label{Ftheta}
\ea
Here, we assume that the two degenerate many-body states of the two-mode problem 
\cite{Bader}, have equal weight
in the ground state, and $\theta$ is their relative phase \cite{Kangsoo}.

The maximal degree of fragmentation (that is when $\theta=0$ as assumed in our numerical 
computations above) becomes   
\bea
{\mathfrak F} =  \frac{4}{21} -\frac34 \left(\frac{6}{7}\right)^{1/5} {G_3^{-4/5}}.
\ea 
Within the validity of the continuum approximation, in the limit of large coupling,
the fragmentation reaches approximately 19\,\%, and is hence significantly lower than in either quasi-1D (80\,\%) and 
quasi-2D (33\,\%) trapping geometries. 
The power law of the asymptotics here is $4/5$, while in quasi-1D trapping  it has been $4/3$ and in quasi-2D unity \cite{Fischer}. 
This implies that the coupling dependence of the degree of fragmentation becomes weaker with increasing dimension.

Finally, we conclude from the comparison with the numerical data shown in Figs.\,\ref{fig:FoverN} and 
\ref{fig:FoverG}, 
that the agreement of two-mode continuum limit and numerics 
is excellent for sufficiently large values of $N$ and $G_3$.

\section{Conclusion}

A dimensionless measure, $G_D$, of the relative importance of total interaction and potential energies, 
which ultimately determines the first-order coherence properties of a trapped system, can be constructed from the
three-dimensional coupling constant $g$ and the relevant trapping lengths in quasi-1D, quasi-2D 
and proper three-dimensional systems.  
The results presented in the above, together with the quasi-1D and quasi-2D counterparts derived 
in \cite{FischerBader}, where we found that the critical $(G_1)_c\sim \ord(10)$ and 
$(G_2)_c\sim \ord(100)$, lead us to conclude that the  dimensionless critical $G_D$ in dimension $D$,  
for trapped dilute Bose gases at absolute zero, scales approximately like $(G_D)_c \sim 10^D$ for fragmentation into two 
macroscopically occupied orbitals to occur. 
We have, furthermore, demonstrated that the degree of fragmentation increases more slowly 
with $G_D$ when the dimension increases. 

The corollary of our result is the asymptotic irrelevance of interactions in large spatial dimensions {($D\geq3$) for the many-body phenomenon of macroscopic fragmentation to occur, due to the exponentially increasing lower bound on the critical interaction strength}.

The correlations leading to fragmented condensate states,  
which force us to go beyond the mean-field theory of a single macroscopically occupied orbital, 
thus become relatively weaker with increasing spatial dimension. 

\acknowledgments
URF was supported by the NRF Korea, grant Nos. 2010-0013103, 2011-0029541, 
and the Seoul National University Foundation Research Expense. 
PB received support from the Ministerio de Ciencia e Innovaci\'on of Spain 
under the projects MTM2009-08587, MTM2010-18246-C03, and the FPU fellowship AP2009-1892.

\end{document}